\documentclass[a4paper,english,aps,manuscript]{revtex4}
\usepackage[T1]{fontenc}
\usepackage[latin9]{inputenc}
\usepackage{prettyref}
\usepackage{amsmath}
\usepackage{amssymb}
\usepackage{esint}

\makeatletter
\@ifundefined{textcolor}{}
{%
 \definecolor{BLACK}{gray}{0}
 \definecolor{WHITE}{gray}{1}
 \definecolor{RED}{rgb}{1,0,0}
 \definecolor{GREEN}{rgb}{0,1,0}
 \definecolor{BLUE}{rgb}{0,0,1}
 \definecolor{CYAN}{cmyk}{1,0,0,0}
 \definecolor{MAGENTA}{cmyk}{0,1,0,0}
 \definecolor{YELLOW}{cmyk}{0,0,1,0}
 }

\makeatother

\makeatother

\usepackage{babel}

\makeatother

\usepackage{babel}

\begin{document}

\title{Generalization of the detailed fluctuation theorem for Non-Hamiltonian
Dynamics}

\author{K. Gururaj, G. Raghavan and M.C. Valsakumar}

\affiliation{Theoretical Studies Section, Material Physics Division, IGCAR, Kalpakkam-603102,
INDIA}

\email{gururaj@igcar.gov.in}
\begin{abstract}
Detailed fluctuation theorem, a microscopic version of the steady
state fluctuation theorem, has been proposed by Jarzynski and demonstrated
in the case of Hamiltonian systems weakly coupled with reservoirs.
We show that an identical theorem for phase space compressibility
rate can be derived for systems evolving under non-Hamiltonian extended
system dynamics, without certain limiting assumptions made in the
original work. Our derivation is based on the non-Hamiltonian phase
space formulation of statistical mechanics and does not rely on any
assumptions of thermodynamic nature. This version of the detailed
fluctuation theorem is shown to be generic enough to be applicable
to several thermostatting schemes. It is shown that in equilibrium,
this detailed fluctuation theorem boils down to the detailed balance
equation and it is further shown to reproduce the Jarzynski's work
theorem for driven systems. 
\end{abstract}
\maketitle

\section{Introduction}

A family of remarkable results in non-equilibrium statistical mechanics,
collectively called \textit{Fluctuation theorems,} has been obtained
in recent years. These theorems address the issue of how macroscopic
irreversibility arises from microscopic time reversible dynamics,
and in a sense quantify the probability of observing Second law violating
events. These fluctuation theorems are valid for systems arbitrarily
far from thermodynamic equilibrium and have been demonstrated for
both deterministic and stochastic evolution. It is this sweeping generality
that make these results quite extraordinary. One version of the fluctuation
theorem was first discovered by Evans and Searles in the context of
molecular dynamics simulation \cite{key-11} of a steady state system
subject to a Gaussian isokinetic thermostat. The system under consideration
was a steady state system subjected to external forcing. Under such
circumstances, stationarity can be attained only under constant dissipation
of energy. In a dynamical system, this dissipation manifests itself
as a contraction of the phase space volume. Considering a set of phase
points centered around $\Gamma_{0}(p_{0},q_{0})$ at time t=0, under
the dynamics, we have the mapping d$\Gamma_{0}$$\rightarrow$d$\Gamma_{t}$
with:

\begin{equation}
\left|\frac{d\Gamma_{0}}{d\Gamma_{t}}\right|=exp(-\intop_{0}^{t}\Lambda(\Gamma_{s})ds)\label{eq:1}\end{equation}
 where \begin{equation}
\Lambda(\Gamma_{s})=\mathit{-\mathit{\mathsf{\frac{\partial}{\partial\Gamma_{s}}\dot{\Gamma_{t}}}}}\label{eq:2}\end{equation}
 then the heat exchange rate with the thermostat, at temperature T,
is \begin{equation}
\dot{Q}(\Gamma,t)=-T\Lambda(\Gamma,t)\label{eq:3}\end{equation}
 which provides the link between phase space compression and entropy
production rate. Evans and Searles then define a dissipation function
:

\begin{equation}
\Omega_{t}(\Gamma_{0})=ln\left[\frac{f(\Gamma_{0,}0)}{f(\Gamma_{t}^{*},0)}\right]-\overset{t}{\underset{0}{\int}}\Lambda(\Gamma_{s},s)ds\label{eq:4}\end{equation}
 where $f(\Gamma_{s},s)$ is the phase space probability density.

For a dissipative system, \begin{equation}
\left\langle \Omega_{t}\right\rangle \geqslant0\label{eq:5}\end{equation}
 When the initial probability density $f(\Gamma_{0},0)$ is drawn
from an equilibrium distribution and the system is driven away from
it by external forcing, the statement of the Evans-Searles (transient)
fluctuation theorem takes the form:

\begin{equation}
\frac{P(\Omega_{t}=A)}{P(\Omega_{t}=-A)}=exp(At)\label{eq:6}\end{equation}
 In the case where the initial phase space points are drawn from a
steady state distribution, Evans and Searles derived a rearranged
form of the above equation called the steady state fluctuation theorem
valid for the long time limit, given by the expression:

\begin{equation}
\underset{t\rightarrow\infty}{lim}\frac{1}{t}ln\frac{P(\overline{\Omega}=A)}{P(\overline{\Omega}=-A)}=A\label{eq:7}\end{equation}
 Independent of this work, Galavotti and Cohen derived the steady
state fluctuation theorem using average compression as:

\begin{equation}
\underset{t\rightarrow\infty}{lim}\frac{1}{t}ln\frac{P(-\bar{\Lambda}=A)}{P(-\bar{\Lambda}=-A)}=A\label{eq:8}\end{equation}
 These fluctuation theorem are valid in the non-linear, nonequilibrium
regime where very general results are available. Fluctuation relation
have been derived under a variety of conditions {[}\cite{key-11}{]}
and have been demonstrated experimentally in small systems.

\section{Jarzynski's detailed fluctuation theorem}

In an interesting development, Jarzynski\textbf{\cite{key-7,key-6}}derived
a hybrid fluctuation theorem\textit{,} which is also a statement of
detailed balance. This \textit{detailed fluctuation theorem} shows
that under a nonequilibrium process, the ratio of the probability
of observing specific trajectory - anti-trajectory pairs goes as an
exponential of the entropy produced. For a trajectory that starts
in a microstate $z_{A}$ and evolves into $z_{B}$ in a duration $\tau$,
resulting in the production of entropy $\triangle S$, its \textit{anti-trajectory}
is the one that starts from $z_{B}^{*}$ and evolves into $z_{A}^{*}$
causing an entropy consumption of $\triangle S$ . Here $(\mathbf{q},\mathbf{p})^{*}$
stands for reversal of momenta $(\mathbf{q},-\mathbf{p})$ . The detailed
fluctuation theorem, like all fluctuation theorems, shows that it
is more likely that entropy is generated rather than consumed in a
non-equilibrium process. This theorem is however distinct from the
other forms of fluctuation theorems in that it makes specific reference
to the initial and final microstates.

Jarzynski has derived this result for a Hamiltonian system weakly
coupled to a set of Hamiltonian reservoirs. The system is manipulated
by an external protocol which involves making or breaking contact
with external forces and heat reservoirs in a specified sequence.
In this scheme, the application of the external protocol in the reversed
order amounts to the realization of the time-reversed trajectories.
A process $\Pi^{+}$ is defined to be the execution of a given protocol
for a time interval $\tau$ and correspondingly, the process $\Pi^{-}$
is the same protocol executed in the reverse order. The detailed fluctuation
theorem then assumes the form\begin{equation}
\frac{P_{+}(\mathbf{z}_{B},\triangle S|\mathbf{z}_{A})}{P_{-}(\mathbf{z}_{A}^{*},-\Delta S\mid\mathbf{z_{\mathbf{B}}^{*}})}=e^{\triangle S/k_{B}}\label{eq:9}\end{equation}
 where the numerator (denominator) in LHS is the joint conditional
probability that, under the process $\Pi^{+}$$(\Pi^{-})$, that the
system starts from $z_{A}$ $(z_{B}^{*})$ and ends at $z_{B}$ $(z_{A}^{*})$,
producing (consuming) an entropy $\triangle S.$ The derivation of
the theorem follows directly from the assumptions that the reservoir
temperatures do not change and their degrees of freedom at both initial
and final times are Maxwell-Boltzmann distributed.The entropy generated
by the dynamical evolution is then assumed to be:

\begin{equation}
\triangle S=-\overset{N}{\underset{n=1}{\sum}}\frac{\Delta Q_{n}}{T_{n}}\label{eq:10}\end{equation}
 where $\Delta Q_{n}$ is the the change in the internal energy of
the n$^{th}$ heat reservoir. These assumptions, coupled with the
fact that Hamiltonian evolution is time reversible and phase-space
conserving, lead directly to the detailed fluctuation theorem. Regarding
the particular choice made for $\triangle S$, eq\prettyref{eq:10},
Jarzynski explicitly states that it is valid only when the system
is weakly coupled with the heat reservoirs possessing infinite number
of degrees of freedom. Towards the end of his paper, Jarzynski suggests
that it would be desirable to perform experiments to test the detailed
fluctuation theorem. This requires experimental control and manipulation
of the full microstates of the system which is very difficult to realize
experimentally. This might explain only very few results are available
on the experimental verification of the detailed fluctuation theorem
in contrast to work fluctuation theorem of Jarzynski. In contrast
to the experimental scenario, in molecular dynamics one has full control
and information about the specific microstate and complete dynamics
of the system and hence one can attempt to verify the detailed fluctuation
theorem in such a setting.

\section{Detailed Fluctuation theorem in Molecular Dynamics}

In the context of molecular dynamics simulations, thermostatting is
realized by appending a few extra degrees of freedom to the system,
the time evolution of which is of a non-Hamiltonian character and
hence cannot be understood as a simple perturbation of the original
uncoupled dynamics. Hence, the question of whether these fluctuation
theorems can be captured in molecular dynamics simulations has been
attracting a lot of attention in recent years. For example, the \textit{work
fluctuation theorem} has been rederived for Nose-Hoover dynamics without
the weak coupling assumption \cite{key-10}. Similarly, the validity
of these fluctuation theorems for different molecular dynamics ensembles
has appeared in literature \cite{key-1,key-2,key-3,key-15}. Within
the context of detailed fluctuation theorem, Jarzynski \cite{key-6,key-7},
has suggested that it would be interesting to see whether the DFT
manifests itself in the Nose-Hoover thermostatted dynamics. Nose-Hoover
dynamics belong to a class of thermostatting schemes called Extended
Phase Space Methods \cite{key-8} and we realize that in the case
of such systems, there are no coupling terms in the total Hamiltonian
and hence one cannot talk of a weak coupling or even delineate the
system energy from the heat bath energy, as was done in the original
derivation. Therefore one cannot use the canonical definition of entropy
(eq 10) . Further, in the extended system dynamics, temperature enters
only as an external parameter and hence taking recourse to thermodynamic
definition of entropy or internal energy is not viable. Given a set
of autonomous equations of motion, perhaps the only quantity that
can be related to entropy is the phase space compressibility. Although
the absolute entropy for such systems may not be definable, one can
identify, under certain assumptions, phase space compression rate
with the entropy production rate\cite{key-14,key-13,key-12}. Independent
of whether such an identification is to be made or not, we demonstrate
that a detailed fluctuation theorem can be formulated for the phase
space compressibility that takes exactly the same form as that of
the original.

\section{Derivation of the detailed fluctuation theorem for the phase compression
rate}

The present derivation follows essentially the same methodology as
that of the original, but differs significantly in that no use is
made of the Liouville theorem since, in general, phase space volume
is not conserved under a non-Hamiltonian evolution. The notations
and symbols used are identical to the ones used in the original, so
as to make the similarities and distinctions between the two derivations
easily observable.

We start by considering a system $\psi$ coupled to a thermostat such
that, the extended system undergoes a time-reversible, non-Hamiltonian
dynamics for which a constant \textit{energy functional} $\mathcal{H}$,
hence-forth referred to as the psuedo-Hamiltonian, can be identified.
It is possible to cast many of the present day thermostats and ergostats
into this form, see \cite{key-9}. Let $\mathbf{z=(}\mathbf{r,p)}$
denote the phase space of the system $\psi$ and let the degrees of
freedom of the thermostat be represented by $\mathbf{Y}$. Let $\mathbf{\Gamma}=(\mathbf{z},\mathbf{Y})$
denote a point in the phase space of the extended system, evolving
under the deterministic non-Hamiltonian dynamics: \cite{key-9}

\begin{equation}
\dot{\mathbf{\mbox{\ensuremath{\Gamma}}}_{i}}=\overset{2Nd}{\underset{j=1}{\sum}}\mathbf{B_{\mathbf{ij}}\frac{\partial\mathcal{H}}{\partial\Gamma_{\mathbf{\mathbf{j}}}}}\label{eq:11}\end{equation}
 for $i=1,...,2Nd$, where $\mathcal{H}$($\Gamma$\textbf{)} is the
generalized energy and \textbf{$B$} is a $2Nd\times2Nd$ anti-symmetric
matrix. The equations are known to generate energy preserving dynamics.
For instance, when $\mathbf{B=\left[\begin{array}{cc}
\mathbf{0} & \mathbf{1}\\
\mathbf{-1} & \mathbf{0}\end{array}\right]}$ where $\mathbf{0}$ and $\mathbf{1}$ indicate the zero and identity
matrices of appropriate dimensions, these equations reduce to the
usual Hamilton's equations of motion. Any other choice of matrix \textbf{$\mathbf{B}$,}
so long as it is anti-symmetric, still generates a non-Hamiltonian
dynamics that conserves \textbf{$\mathcal{H}$:}

\begin{equation}
\frac{d\mathcal{H}}{dt}=\overset{2Nd}{\underset{i=1}{\sum}}\overset{2Nd}{\;\underset{j=1}{\sum}}\frac{\partial\mathcal{H}}{\partial\Gamma_{i}}\mathbf{B}_{ij}\frac{\partial\mathcal{H}}{\partial\Gamma_{j}}=0\label{eq:12}\end{equation}
 We further note that time reversibility of the underlying dynamics
enforces the condition

\begin{center}
\begin{equation}
\mathcal{H}(\Gamma)=\mathcal{H}(\Gamma^{*})\label{eq:13}\end{equation}

\par\end{center}

\begin{flushleft}
where $\Gamma^{*}$ is obtained from $\Gamma$ by reversing all the
momentum-like variables. The phase space contraction rate $\Lambda(\Gamma^{s})$
at time $t=s$, defined earlier, may be written as 
\par\end{flushleft}

\begin{equation}
\Lambda(\Gamma)_{t=s}=-\underset{i=1}{\overset{2Nd}{\sum}}\dfrac{\partial\dot{\Gamma}_{i}(t)}{\partial\Gamma_{i}(t)}\mid_{t=s}=-\overset{2Nd}{\underset{i,j=1}{\sum}}\frac{\partial\mathbf{B}_{ij}}{\partial\Gamma_{i}}\frac{\partial\mathcal{H}}{\partial\Gamma_{j}}\mid_{t=s}\label{eq:14}\end{equation}
 We now define entropy production in a time interval $\tau$ under
a non-Hamiltonian evolution be\begin{equation}
\Delta S=k_{B}\underset{0}{\overset{\tau}{\int}}\Lambda(\Gamma^{t})dt\label{eq:15}\end{equation}
 where $k_{B}$ denotes the Boltzmann constant. For a given state
$\mathbf{\Gamma}=(\mathbf{z},\mathbf{Y})$, we define a time reversed
state $\mathbf{\Gamma^{*}}$ to be $(\mathbf{z}^{*},\mathbf{Y}^{*})$
. Let $\Gamma_{+}(t)$ denote the state of the extended system at
time $t$ starting from an initial extended microstate $\mathbf{\Gamma}^{0}$
and let $\Gamma_{-}(t)$ denote state of the system at time $t$ but
evolving from the momentum reversed final state $\mathbf{\Gamma}^{\tau*}$.
With this as the background, we are interested in computing the probability
that the system $\psi$, at $t=0$, starts in a particular microstate
$\mathbf{\mathbf{z}}(0)=\mathbf{z}_{A}$ and after evolving for a
time $t=\tau$, reaches a state $\mathbf{\mathbf{z}}(\tau)=\mathbf{z}_{B}$,
generating an entropy $\triangle S$. Note that the evolution of $\mathbf{\Gamma}=(\mathbf{z},\mathbf{Y})$
is deterministic, eq\prettyref{eq:11}. We resort to probabilistic
description when we are focusing only on the evolution of $\mathbf{z}$.
For a particular choice of the initial condition $\mathbf{z}=\mathbf{z}_{A}$,
the probability distribution is over the initial conditions of heatbath
variables \textbf{Y}\begin{equation}
P_{+}(\mathbf{z}_{B},\triangle S|\mathbf{z}_{A})=\int d\mathbf{Y}p(\mathbf{Y})\delta[\mathbf{z}_{B}-\mathbf{z_{+}^{\tau}(}\mathbf{z_{A},}\mathbf{Y)].}\delta[\triangle S-\hat{\triangle S}(\mathbf{z_{A},}\mathbf{Y})]\label{eq:16}\end{equation}
 where $\mathbf{z_{+}^{\tau}(}\mathbf{z_{A},}\mathbf{Y)}$ and $\hat{\triangle S}(\mathbf{z_{A},}\mathbf{Y})$
denote the final state of the system $\psi$ and the net entropy produced
respectively, after $\psi$ has evolved for a time $t=\tau$ starting
at $\mathbf{z_{A}}$ with the reservoir starting at $\mathbf{Y}.$
Using the identity $\mathbf{z_{A}}=\mathbf{\hat{z}_{+}^{0}(}\mathbf{z,}\mathbf{Y)}$
we can now integrate over the full phase space vector $\mathbf{\Gamma}=(\mathbf{z},\mathbf{Y})$:\begin{equation}
P_{+}(\mathbf{z}_{B},\triangle S|\mathbf{z}_{A})=\int d\mathbf{\Gamma}p(\mathbf{\Gamma})\delta[\mathbf{z_{A}}-\mathbf{\hat{z}_{+}^{0}(}\mathbf{\Gamma)}].\delta[\mathbf{z}_{B}-\mathbf{z_{+}^{\tau}(}\mathbf{\Gamma})].\delta[\triangle S-\hat{\triangle S}(\mathbf{\Gamma})]\label{eq:17}\end{equation}
 if the dynamical system eq \prettyref{eq:11} possess $n_{c}$ independent
conservation laws:

\begin{flushleft}
\begin{equation}
\Lambda_{k}(\Gamma)=\lambda_{k},\quad k=1,..,n_{c}\label{eq:18}\end{equation}
 then the probability distribution of the mirostates can be written
as: \cite{key-8} 
\par\end{flushleft}

\begin{equation}
p(\Gamma)=\overset{n_{c}}{\underset{k=1}{\prod}}\delta(\Lambda_{k}(\Gamma)-\lambda_{k})\label{eq:19}\end{equation}
 In instances where $\mathcal{H}$ is the only conserved quantity
of the dynamics, the states $\mathbf{\Gamma}$ are those drawn from
the constant $\mathcal{H}$ surface

\begin{equation}
p(\Gamma^{0})=\delta(H(\mathbf{\Gamma^{0}})-E)\label{eq:20}\end{equation}
 hence, eq \prettyref{eq:17} may be rewritten as

\begin{flushleft}
\begin{equation}
P_{+}(\mathbf{z}_{B},\triangle S|\mathbf{z}_{A})=\int d\mathbf{\Gamma^{0}}\delta(\mathcal{H}(\Gamma^{0})-E)\delta[\mathbf{z_{A}}-\mathbf{\hat{z}^{0}(}\mathbf{\Gamma^{0})}].\delta[\mathbf{z}_{B}-\mathbf{z^{\tau}(}\Gamma^{0})].\delta[\triangle S-\hat{\triangle S}(\Gamma^{0})]\label{eq:21}\end{equation}
 
\par\end{flushleft}

\begin{flushleft}
As the dynamics is time reversal invariant, and $\mathcal{H}$ is
conserved and we have: 
\par\end{flushleft}

\begin{equation}
\triangle\hat{S}(\Gamma^{\tau*})=-\hat{\triangle S}(\Gamma^{0})\label{eq:22}\end{equation}
 \begin{equation}
\mathcal{H}(\mathbf{\Gamma^{0}})=\mathcal{H}(\mathbf{\Gamma}^{\tau*})\label{eq:23}\end{equation}
 \begin{equation}
\mathbf{\hat{z}^{0}(}\mathbf{\Gamma^{0})}=\left[\mathbf{z^{\tau}(}\Gamma^{\tau*})\right]^{*}\label{eq:24}\end{equation}
 With these expressions at hand, we may now recast the above integral
as:

\begin{equation}
P_{+}(\mathbf{z}_{B},\triangle S|\mathbf{z}_{A})=\int d\mathbf{\Gamma^{0}}\delta(\mathcal{H}(\mathbf{\Gamma}^{\tau*})-E)\delta[\mathbf{z_{A}}^{*}-\mathbf{\hat{z}^{\tau}(}\mathbf{\mathbf{\Gamma}^{\tau*})}].\delta[\mathbf{z}_{B}^{*}-\mathbf{\hat{z}^{\mathbf{0}}(}\mathbf{\mathbf{\Gamma}^{\tau*}})].\delta[\triangle S+\hat{\triangle S}(\mathbf{\Gamma}^{\tau*})]\label{eq:25}\end{equation}
 where we have cast all the variables in the r.h.s of the above integral
with the time reversed counterparts of the final states, except for
the integration volume element $d\Gamma^{0}$. In general, as the
phase space volume is not conserved in a non-Hamiltonian evolution,
$d\Gamma^{0}$ is not an invariant volume under eq\prettyref{eq:11}
and hence $d\Gamma^{0}$cannot be directly replaced by $d\Gamma^{\tau*}$.
The Jacobian of transformation from the initial to the final time
reversed phase space coordinates $J(\mathbf{\Gamma}^{\tau*};\mathbf{\Gamma}^{0})$
is not unity and we have

\begin{equation}
d\mathbf{\Gamma}^{\tau*}=J(\mathbf{\Gamma}^{\tau*};\mathbf{\Gamma}^{0})d\mathbf{\Gamma}^{0}=J(\mathbf{\Gamma}^{\tau*};\mathbf{\Gamma}^{\tau})J(\mathbf{\Gamma}^{\tau};\mathbf{\Gamma}^{0})d\Gamma^{0}\label{eq:26}\end{equation}
 As the dynamical system eq\prettyref{eq:11} is assumed to be time
reversal invariant, \begin{equation}
J(\mathbf{\Gamma}^{\tau*};\mathbf{\Gamma}^{\tau})=1\label{eq:27}\end{equation}
 therefore, \begin{equation}
J(\mathbf{\Gamma}^{\tau*};\mathbf{\Gamma}^{0})=J(\mathbf{\Gamma}^{\tau};\mathbf{\Gamma}^{0})\label{eq:28}\end{equation}
 The time evolution of Jacobian of transformation is: \cite{key-8}

\begin{equation}
\dfrac{dJ}{dt}=J\underset{l=1}{\overset{2Nd}{\sum}}\dfrac{\partial\dot{\Gamma_{l}}(t)}{\partial\Gamma_{l}(t)}\label{eq:29}\end{equation}
 The sum on the right is nothing but the negative of the phase space
compressibility rate $\Lambda(\Gamma^{t})$ , hence:

\begin{equation}
J(\Gamma^{\tau};\Gamma^{0})=e^{\underset{0}{\overset{\tau}{-\int}}\Lambda(\Gamma^{t})dt}=e^{-\frac{\hat{\triangle S(}\mathbf{\Gamma}^{0})}{k_{B}}}\label{eq:30}\end{equation}
 From eq\prettyref{eq:22}, we know that $\triangle\hat{S}(\Gamma^{\tau*})=-\hat{\triangle S}(\Gamma^{0})$
and we have,

\begin{equation}
d\mathbf{\Gamma}^{\tau*}=e^{\frac{\hat{\triangle S}(\mathbf{\Gamma}^{\tau^{*}})}{k_{B}}}d\mathbf{\Gamma}^{0}\label{eq:31}\end{equation}
 We may now replace $d\mathbf{\Gamma}^{0}$ in eq\prettyref{eq:25}
with $e^{-\frac{\hat{\triangle S}(\mathbf{\Gamma}^{\tau^{*}})}{k_{B}}}d\mathbf{\Gamma}^{\tau*}$
and we have:

\begin{equation}
P_{+}(\mathbf{z}_{B},\triangle S|\mathbf{z}_{A})=\int e^{-\frac{\hat{\triangle S}(\mathbf{\Gamma}^{\tau^{*}})}{k_{B}}}d\mathbf{\mathbf{\Gamma}^{\tau*}}\delta(\mathcal{H}(\mathbf{\Gamma}^{\tau*})-E)\delta[\mathbf{z_{A}}-\mathbf{\hat{z}^{0}(}\mathbf{\mathbf{\Gamma}^{\tau*})}].\delta[\mathbf{z}_{B}-\mathbf{z^{\tau}(}\mathbf{\mathbf{\Gamma}^{\tau*}})].\delta[\triangle S+\hat{\triangle S}(\mathbf{\Gamma}^{\tau*})],\label{eq:32}\end{equation}
 We can pull out the factor $e^{\frac{\triangle S}{k_{B}}}$ out of
the integration due to the presence of the delta function $\delta[\triangle S+\hat{\triangle S}(\mathbf{\Gamma}^{\tau*})]$.

\begin{equation}
P_{+}(\mathbf{z}_{B},\triangle S|\mathbf{z}_{A})=e^{\frac{\triangle S}{k_{B}}}\int d\mathbf{\mathbf{\Gamma}^{\tau*}}\delta(\mathcal{H}(\mathbf{\Gamma}^{\tau*})-E)\delta[\mathbf{z_{A}}-\mathbf{\hat{z}^{0}(}\mathbf{\mathbf{\Gamma}^{\tau*})}].\delta[\mathbf{z}_{B}-\mathbf{z^{\tau}(}\mathbf{\mathbf{\Gamma}^{\tau*}})].\delta[\triangle S+\hat{\triangle S}(\mathbf{\Gamma}^{\tau*})],\label{eq:33}\end{equation}
 The right hand side of the above equation is nothing but $e^{\frac{\Delta S}{k_{B}}}$
times the probability distribution $P(\mathbf{z}_{B}^{*},-\triangle S|\mathbf{z}_{A}^{*})$
. Thus we have the final result of the Detailed Fluctuation theorem:

\begin{flushleft}
\begin{equation}
P(\mathbf{z}_{B},\triangle S|z_{A})=e^{\frac{\Delta S}{k_{B}}}P(\mathbf{z}_{B}^{*},-\triangle S|\mathbf{z}_{A}^{*}),\label{eq:34}\end{equation}
 In summary, from the ingredients of the above derivation we list
all the requirements that a dynamical system has to satisfy for the
present derivation to go through: 
\par\end{flushleft}

1. The phase space of the extended system is of even number of dimensions.

2. The evolution is deterministic and equations of motion are time
reversal invariant.

3. The phase space compressibility of the system is non-zero.

4. A constant of motion, usually a {}``psuedo-Hamiltonian'' can
be identified for the system that is time reversal invariant.

From the above points, it can be see that no assumptions have been
made about the specific type of interactions present in the system
and there are no restrictions on system size and time scale $\tau$
over which these fluctuations are realized. It should also be appreciated
that this detailed fluctuation theorem is valid arbitrarily far from
equilibrium, as no assumptions of thermodynamic nature has been made
in the derivation. Further, unlike in the original derivation, there
is no requirement for the system to be driven out of equilibrium through
a given protocol. As there are fluctuations in the phase space compressibility
even at equilibrium, one can capture the detailed fluctuation theorem
even in an equilibrium simulation. This result is in contrast to many
other fluctuation theorems which are valid far from equilibrium and
at equilibrium they boil down to trivial identities as there is no
average entropy production or consumption at equilibrium.

Another issue is the choice of the microcanonical ensemble for the
extended system. This is in contrast with Jarzynski's suggestion,
which is to sample the reservoir degree of freedom from a Gaussian
distribution. Instead, the present derivation treats all the degrees
of freedom on equal footing and hence is more appealing, where the
only assumptions made are the usual ones of ergodicity and equal apriori
probabilities\cite{key-8,key-10}.

\section{Illustration of detailed fluctuation theorem in Different ensembles}

For a better appreciation of our results, we shall study the detailed
fluctuation theorem in the context of a few popular extended phase
space methods. All the examples mentioned below have the non-Hamiltonian
evolution such that the system of interest evolves consistent with
the appropriate ensemble the equations are supposed to mimic. The
statistical mechanical properties of these systems are very extensively
studied in the earlier papers {[}\cite{key-16,key-8,key-18}{]} and
hence our interest here will be limited to examining them in the light
of the applicability of the detailed fluctuation theorem.

\subsection{NOSE HOOVER DYNAMICS}

In the Nose-Hoover thermostatting scheme \cite{key-8}, an N particle
system in $d$ spatial dimensions with Cartesian positions $\mathbf{r}\equiv\{\mathbf{r_{1}},...,\mathbf{r_{N}}\mathbf{\}}$
and momenta $\mathbf{p\equiv\{p}_{1},...,\mathbf{p_{N}\}}$ interacting
through N particle potential $\Phi(\mathbf{r}_{1},...,\mathbf{r}_{N})$
is augmented with 2 heat bath variables $\mathbf{Y=}(\eta,p_{\eta})$
such that the $2dN+2$ dimensional extended phase space vector $\mathbf{\Gamma}=(\mathbf{r},\mathbf{p},\eta,p_{\eta})$
evolves as follows:

\begin{equation}
\mathbf{\dot{r}_{i}=}\dfrac{\mathbf{p_{i}}}{m}\label{eq:35}\end{equation}

\begin{equation}
\mathbf{\dot{p}_{i}}=-\dfrac{\partial\Phi(\{\mathbf{r_{i}\})}}{\partial\mathbf{r_{i}}}-\dfrac{p_{\eta}}{Q}\mathbf{p_{i}}\label{eq:36}\end{equation}

\begin{equation}
\dot{\eta}=\dfrac{p_{n}}{Q}\label{eq:37}\end{equation}
 \begin{equation}
\dot{p_{\eta}}=\left[\overset{N}{\underset{i=1}{\sum}}\dfrac{\mathbf{p}_{i}^{2}}{m}-dNk_{B}T\right]\label{eq:38}\end{equation}

The parameter $Q$ represents the strength of the Nose-Hoover coupling
which controls the time scale over which the equilibration takes place
and $T$ is the temperature at which we wish to maintain the system
of interest. A curious point that can be noted in these equations
is that the variable $\eta$ does not explicitly get connected to
other degrees of freedom. Still, it is profitable to retain this variable
in the dynamical equations of motion as it facilitates the casting
of these equations in the desired form \prettyref{eq:11}

\begin{equation}
\left(\begin{array}{c}
\dot{\mathbf{r}_{i}}\\
\dot{\eta}\\
\mathbf{\dot{p_{i}}}\\
\dot{p_{\eta}}\end{array}\right)=\left[\begin{array}{cccc}
0 & 0 & 1 & 0\\
0 & 0 & 0 & 1\\
-1 & 0 & 0 & -\mathbf{p_{i}}\\
0 & -1 & \mathbf{p_{i}} & 0\end{array}\right]\left(\begin{array}{c}
\frac{\partial\Phi(\{\mathbf{r_{i}}\})}{\partial\mathbf{r_{i}}}\\
dNk_{B}T\\
\frac{\mathbf{p}_{i}}{m}\\
\frac{p_{\eta}}{Q}\end{array}\right)\label{eq:39}\end{equation}
 for\begin{equation}
\mathcal{H}(\Gamma)=\overset{N}{\underset{i=1}{\sum}}\dfrac{\mathbf{p}_{i}^{2}}{2m_{i}}+\Phi(\{\mathbf{r\}})+\dfrac{p_{\eta}^{2}}{2Q}+dNk_{B}T\eta\label{eq:40}\end{equation}
 Identifying the first two terms of the above as the system Hamiltonian,

\begin{equation}
H^{s}(\mathbf{\{r}\},\{\mathbf{p\}})=\overset{N}{\underset{i=1}{\sum}}\dfrac{\mathbf{p}_{i}^{2}}{2m_{i}}+\Phi(\{\mathbf{r\}})\label{eq:41}\end{equation}
 we have \begin{equation}
\mathcal{H}(\Gamma)=H^{s}(\{\mathbf{r\}},\{\mathbf{p\}})+\dfrac{p_{\eta}^{2}}{2Q}+dNk_{B}T\eta\label{eq:42}\end{equation}
 Nose-Hoover dynamics is time reversal invariant and possess a {}``psuedo-Hamiltonian''
as a constant of motion and hence, satisfies the essential requirements
needed for the present analysis. The phase space compressibility of
this system is given by eq \prettyref{eq:14}

\begin{equation}
\Lambda(\Gamma^{s})=-\overset{2n+2}{\underset{i=1}{\sum}}\frac{\partial\dot{\Gamma^{i}}}{\partial\Gamma^{i}}=-\underset{i}{\sum}\frac{\partial\mathbf{p_{i}}}{\partial\mathbf{p}_{i}}=\dfrac{dNp_{\eta}}{Q}=dN\dot{\eta}\label{eq:43}\end{equation}
 so the total phase space compression during an evolution from t=0
to t=$\tau$ is from eq \prettyref{eq:15}

\begin{equation}
\Delta S=k_{B}\underset{0}{\overset{\tau}{\int}}\Lambda(\Gamma^{s})ds=k_{B}dN(\eta(\tau)-\eta(0))\label{eq:44}\end{equation}
 There are several points in order. Given a time evolution of $\Gamma^{0}=(\mathbf{r_{i}^{0},p_{i}^{0}},\eta^{0},p_{\eta}^{0})$
to $\Gamma^{t}=(\mathbf{r_{i}^{t},p_{i}^{t}},\eta^{t},p_{\eta}^{t})$,
there is a phase space compression of $k_{B}dN(\eta^{t}-\eta^{0})$.
Note that this quantity is path independent and depends only on the
initial and final state of the position variable of the heat bath,
and does not explicitly depend upon the functional form of the potential.

With this definition, one can gather all the trajectories that start
at $(\mathbf{z_{A},}\eta,p_{\eta})$ and evolve to $(\mathbf{z_{B},}\eta+\triangle S,\acute{p}_{\eta})$
where $\mathbf{z_{A}}$and $\mathbf{z_{B}}$ are fixed initial and
final states (positions and momenta) of the system of interest and
$\eta$,$p_{\eta}$and $\acute{p_{\eta}}$ are arbitrary (subject
to the requirement that the extended state vector $\Gamma$ stays
on a constant energy hypersurface). The probability of obtaining such
a trajectory is $P_{+}(\mathbf{z_{B}},+\triangle S|\mathbf{z_{A})}$.
To obtain the anti-trajectories, from the time reversed final states
one gathers all those states that start from $(\mathbf{z_{B}^{*},}\eta,\acute{p}_{\eta})$
and evolve into $(\mathbf{z_{A}^{*},}\eta-\triangle S,p_{\eta})$.
The probability of obtaining such a trajectory is $P_{-}(\mathbf{z_{A}^{*}},-\triangle S|\mathbf{z_{B}^{*})}$.
The detailed fluctuation theorem requires the ratio of these two probabilities
go like $e^{\frac{\triangle S}{k_{B}}}$.

Further, in the limit of $Q\rightarrow\infty$, the equations of motion
eq\prettyref{eq:35}and eq\prettyref{eq:36} are reduced to that of
a Hamiltonian evolution and we have, from eq\prettyref{eq:37}, $\eta(\tau)=\eta(0)$
and hence, from eq \prettyref{eq:44}, $\Delta S=0$. This means that
the probabilities of forward and backward trajectories are identical:

\begin{equation}
P(\mathbf{z}_{A}\rightarrowtail\mathbf{z_{B}})=P(\mathbf{z}_{B}^{*}\mathbf{\rightarrowtail z_{A}^{*}})\label{eq:45}\end{equation}
 as can be expected for an isolated system.

Now, consider the probability that the system $\psi$ evolves from
state $\mathbf{z_{A}}$to state $\mathbf{z_{B}}$ in a time $\tau$.
This probability, let it be denoted by $P(\mathbf{z}_{A}\rightarrowtail\mathbf{z_{B}})$
, is

\begin{equation}
P(\mathbf{z}_{A}\rightarrowtail\mathbf{z_{B}})=\int P(\mathbf{z}_{B},\triangle S|\mathbf{z}_{A})d(\Delta S)\label{eq:46}\end{equation}
 Now, applying the detailed fluctuation theorem on the right hand
side, we have

\begin{equation}
P(\mathbf{z}_{A}\rightarrowtail\mathbf{z_{B}})=\int e^{\frac{\Delta S}{k_{B}}}P(\mathbf{z}_{A}^{*},-\triangle S|\mathbf{z}_{B}^{*})d(\Delta S)\label{eq:47}\end{equation}
 For the Nose-Hoover thermostat, we know that $\triangle S=k_{B}dN(\eta(\tau)-\eta(0))$
and from eq\prettyref{eq:42} we have

\begin{equation}
\eta(0)=\frac{\left[E-H^{s}(\mathbf{z_{A}})-\dfrac{p_{\eta}^{2}(0)}{2Q}\right]}{dNk_{B}T},\qquad\qquad\eta(\tau)=\frac{\left[E-H^{s}(\mathbf{z_{B}})-\dfrac{p_{\eta}^{2}(\tau)}{2Q}\right]}{dNk_{B}T}\label{eq:48}\end{equation}
 where E is the constant energy over which the microcanonical ensemble
of the full system is defined. Hence we have,

\begin{equation}
\frac{\Delta S}{k_{B}}=dN(\eta(\tau)-\eta(0))=\frac{H^{s}(\mathbf{z_{A}})-\dfrac{p_{\eta}^{2}(0)}{2Q}-H^{s}(\mathbf{z_{B}})+\dfrac{p_{\eta}^{2}(\tau)}{2Q}}{k_{B}T}\label{eq:49}\end{equation}
 Inserting this into the right hand side of eq\prettyref{eq:47},

\begin{equation}
P(\mathbf{z}_{A}\rightarrowtail\mathbf{z_{B}})=\int e^{\frac{H^{s}(\mathbf{z_{A}})-\dfrac{p_{\eta}^{2}(0)}{2Q}-H^{s}(\mathbf{z_{B}})+\dfrac{p_{\eta}^{2}(\tau)}{2Q}}{k_{B}T}}P(\mathbf{z}_{A}^{*},-\triangle S|\mathbf{z}_{B}^{*})d(\Delta S)\label{eq:50}\end{equation}
 Since the $\mathbf{z}_{A}$ and $\mathbf{z}_{B}$ are independent
of the integration variable, they can be pulled out of the integration
and we have

\begin{equation}
P(\mathbf{z}_{A}\rightarrowtail\mathbf{z_{B}})=e^{\frac{H^{s}(\mathbf{z_{A}})-H^{s}(\mathbf{z_{B}})}{k_{B}T}}\int e^{\frac{p_{\eta}^{2}(\tau)-p_{\eta}^{2}(0)}{2Qk_{B}T}}P(\mathbf{z}_{A}^{*},-\triangle S|\mathbf{z}_{B}^{*})d(\Delta S)\label{eq:51}\end{equation}
 Re-arranging the above,

\begin{equation}
e^{\frac{-H^{s}(\mathbf{z_{A}})}{k_{B}T}}P(\mathbf{z}_{A}\rightarrowtail\mathbf{z_{B}})=e^{\frac{-H^{s}(\mathbf{z_{B}})}{k_{B}T}}\int e^{\frac{p_{\eta}^{2}(\tau)-p_{\eta}^{2}(0)}{2Qk_{B}T}}P(\mathbf{z}_{A}^{*},-\triangle S|\mathbf{z}_{B}^{*})d(\Delta S)\label{eq:52}\end{equation}
 Assuming that the times $t=0$ and $t=\tau$ are chosen sufficiently
long time after the system has equilibrated, we can assume that the
kinetic energy distribution is consistent with the equipartition theorem,
and hence the total kinetic energy is equal to the $\frac{1}{2}dNk_{B}T$.

\begin{equation}
\overset{N}{\underset{i=1}{\sum}}\dfrac{\mathbf{p}_{i}^{2}}{2m}=\frac{1}{2}dNk_{B}T\label{eq:53}\end{equation}
 With this assumption, and from the last of the Nose-Hoover equations
eq\prettyref{eq:38}, we see that $\dot{p_{\eta}}=0$, and hence $e^{\frac{p_{\eta}^{2}(\tau)-p_{\eta}^{2}(0)}{2Qk_{B}T}}=1.$
Substituting this in eq \prettyref{eq:52}, we get

\begin{equation}
e^{\frac{-H^{s}(\mathbf{z_{A}})}{k_{B}T}}P(\mathbf{z}_{A}\rightarrowtail\mathbf{z_{B}})=e^{\frac{-H^{s}(\mathbf{z_{B}})}{k_{B}T}}\int P(\mathbf{z}_{A}^{*},-\triangle S|\mathbf{z}_{B}^{*})d(\triangle S)\label{eq:54}\end{equation}
 The integral in right hand side of the above equation can readily
be identified as the probability $P(\mathbf{z}_{B}^{*}\rightarrowtail\mathbf{z_{A}^{*}})$
and since the system Hamiltonian also the time reversal invariance,
this probability is also equal to $P(\mathbf{z}_{B}\rightarrowtail\mathbf{z_{A}}).$
So, we have:

\begin{equation}
e^{\frac{-H^{s}(\mathbf{z_{A}})}{k_{B}T}}P(\mathbf{z}_{A}\rightarrowtail\mathbf{z_{B}})=e^{\frac{-H^{s}(\mathbf{z_{B}})}{k_{B}T}}P(\mathbf{z}_{B}\rightarrowtail\mathbf{z_{A}})\label{eq:55}\end{equation}
 Which is nothing but the statement of detailed balance, which is
valid for any system at equilibrium.

\subsection{NOSE HOOVER CHAIN DYNAMICS}

If more than one conservation laws are obeyed by the dynamical system,
it is well known that the Nose-Hoover thermostatting scheme fails
to generate the canonical ensemble. This happens because the accessible
phase space gets restricted by the conservation and hence the system
fails to access all the regions of the phase space in the course of
its dynamics. This problem can be overcome by extending the number
of heat bath variables. One such method is the Nose-Hoover chain thermostat.
Its equations of motion are given by:

\begin{equation}
\mathbf{\dot{r}_{i}=}\dfrac{\mathbf{p_{i}}}{m}\label{eq:56}\end{equation}
 \begin{equation}
\mathbf{\dot{p}_{i}}=-\dfrac{\partial\Phi(\{\mathbf{r_{i}\}})}{\partial\mathbf{r_{i}}}-\dfrac{p_{\eta_{1}}}{Q_{1}}\mathbf{p_{i}}\label{eq:57}\end{equation}

\begin{equation}
\dot{\eta_{j}}=\frac{p_{\eta_{j}}}{Q_{j}},\quad j=1,...,M\label{eq:58}\end{equation}
 \begin{equation}
\dot{p_{\eta_{1}}}=\left[\overset{N}{\underset{i=1}{\sum}}\dfrac{\mathbf{p}_{i}^{2}}{m_{i}}-dNk_{B}T\right]-\frac{p_{\eta_{2}}}{Q_{2}}p_{\eta_{1}}\label{eq:59}\end{equation}

\begin{center}
\begin{equation}
\dot{p_{\eta_{j}}}=\left[\dfrac{p_{\eta_{j-1}}^{2}}{Q_{j-1}}-k_{B}T\right]-\frac{p_{\eta_{j+1}}}{Q_{j+1}}p_{\eta_{j}}\quad j=2,...,M-1\label{eq:60}\end{equation}

\par\end{center}

\begin{flushleft}
\begin{equation}
\dot{p_{\eta_{M}}}=\left[\dfrac{p_{\eta_{M-1}}^{2}}{Q_{M-1}}-k_{B}T\right]\label{eq:61}\end{equation}
 It can be shown that these equations of motion, preserve the psuedo-Hamiltonian: 
\par\end{flushleft}

\begin{equation}
\mathcal{H}(\Gamma)=\overset{N}{\underset{i=1}{\sum}}\dfrac{\mathbf{p}_{i}^{2}}{2m_{i}}+\Phi(\{\mathbf{r\}})+\overset{N}{\underset{j=1}{\sum}}\dfrac{p_{\eta_{j}}^{2}}{2Q_{j}}+dNkT\eta_{1}+k_{B}T(\overset{M}{\underset{k=2}{\sum}}\eta_{k})\label{eq:62}\end{equation}
 The phase space compressibility for this system of equations is given
by

\begin{equation}
\Lambda(\Gamma^{s})=(dN\dot{\eta}_{1}+\dot{\eta}_{2}+\dot{\eta}_{3}+...+\dot{\eta}_{M})\label{eq:63}\end{equation}
 Again, the phase space compressibility rate for this system is also
a total time derivative of the heat bath variables and the psuedo-Hamiltonian
is invariant under time reversal. Hence we can derive the detailed
fluctuation theorem in the context of Nose-Hoover chain thermostat.

\subsection{MTK ISOBARIC ENSEMBLE}

The Martyna, Tobias and Klein ensemble is also based on the Non-Hamiltonian
phase space formulation and is known to generate the correct isobaric
ensemble. The equations of motion read (\cite{key-18})

\begin{equation}
\mathbf{\dot{r}_{i}=}\dfrac{\mathbf{p_{i}}}{m}+\frac{\mathbf{p}_{g}}{W_{g}}\mathbf{r_{i}}\label{eq:64}\end{equation}
 \begin{equation}
\mathbf{\dot{p}_{i}}=-\dfrac{\partial\Phi(\{\mathbf{r_{i}\})}}{\partial\mathbf{r_{i}}}-\frac{\mathbf{p}_{g}}{W_{g}}\mathbf{p_{i}}-\frac{1}{N_{f}}\frac{Tr(\mathbf{p}_{g})}{W_{g}}\mathbf{p_{i}}-\frac{p_{\eta_{1}}}{Q_{1}}\mathbf{p_{i}}\label{eq:65}\end{equation}

\begin{equation}
\mathbf{\dot{h}=}\frac{\mathbf{p_{g}h}}{W_{g}}\label{eq:66}\end{equation}
 \begin{equation}
\mathbf{\dot{p}_{g}=}det[\mathbf{h}](\mathbf{P^{(int)}-I}P)+\frac{1}{N_{f}}\sum\mathbf{\frac{p_{i}^{2}}{\mathbf{m}}I}-\frac{p_{\xi_{1}}}{\acute{Q}_{1}}\mathbf{p}_{g}\label{eq:67}\end{equation}

\begin{equation}
\dot{\eta_{j}}=\frac{p_{\eta_{j}}}{Q_{j}}\qquad\dot{\xi}_{j}=\frac{p_{\xi_{j}}}{\acute{Q_{j}}}\label{eq:68}\end{equation}

\begin{equation}
\dot{p_{\eta_{M}}}=G_{M}\label{eq:69}\end{equation}

\begin{equation}
\dot{p_{\xi_{j}}}=\acute{G}_{j}-\frac{p_{\xi_{j+1}}}{\acute{Q_{j+1}}}p_{\xi_{j}}\qquad\dot{p}_{\xi_{M}}=\acute{G}_{M}\label{eq:70}\end{equation}

where \textbf{$\mathbf{P}^{(int)}$}is the internal pressure, \textbf{$I$
}is the 3x3 identity matrix and $G_{j}$are the thermostat forces,
given by

\begin{equation}
G_{1}=\overset{N}{\underset{i=1}{\sum}}\dfrac{\mathbf{p}_{i}^{2}}{m_{i}}-dk_{B}T\qquad G_{j}=\frac{p_{\eta_{j-1}}^{2}}{Q_{j-1}}-k_{B}T\label{eq:71}\end{equation}

and

\begin{equation}
\acute{G}_{1}=\frac{Tr[\mathbf{p_{g}^{T}p_{g}]}}{W_{g}}-d^{2}kT\qquad\acute{G}_{j}=\frac{p_{\xi_{j-1}}^{2}}{Q_{j-1}}-k_{B}T\label{eq:72}\end{equation}

These equations have the psuedo-Hamiltonian as

\begin{equation}
H=\underset{i}{\sum}\dfrac{\mathbf{p}_{i}^{2}}{2m_{i}}+\Phi(\{\mathbf{r\}})+\frac{Tr(\mathbf{p_{g}^{T}p_{\mathbf{g}}})}{2W_{g}}+Pdet(\mathbf{h})+\underset{j=2}{\overset{M}{\sum}}\left[\dfrac{p_{\eta_{j}}^{2}}{2Q_{j}}+\dfrac{p_{\xi_{j}}^{2}}{2\acute{Q_{j}}}\right]+N_{f}k_{B}T\eta_{1}+d^{2}k_{B}T\xi_{1}+k_{B}T\underset{k=2}{\overset{M}{\sum}}(\eta_{k}+\xi_{k})\label{eq:73}\end{equation}
 The phase space compression rate for this system of equations comes
out to be

\begin{equation}
\Lambda=(d-1)\frac{d}{dt}ln[det(\mathbf{h})]+dN\dot{\eta}_{1}+d^{2}\dot{\xi_{1}}+\underset{k=2}{\overset{M}{\sum}}\left[\dot{\eta_{k}}+\dot{\xi_{k}}\right]\label{eq:74}\end{equation}
 It is readily evident that, as with Nose-Hoover, Nose-Hoover Chain
and massive thermostatting schemes, the phase space compression rate
is again a total time derivative of the heat bath variables alone
and the psuedo-Hamiltonian is invariant under time reversal operation.
Further, the phase space compressibility is a function of position-like
variable of the heat bath and hence satisfies the assumption, eq\prettyref{eq:22}.
This implies that the MTK isobaric ensemble is capable of capturing
the detailed fluctuation theorem.

Also, from eq \prettyref{eq:43}, eq \prettyref{eq:63} and eq\prettyref{eq:74},
we see that the phase space compression for all the extended system
dynamics is dependent only on position-like variables of the heat
bath and not on the system variables per se. This is evident from
the extended phase space formulation itself where there are no coupling
terms between the heat bath variables and our Hamiltonian system variables.
The phase space compression for the Hamiltonian systems is always
zero, hence the contribution towards the phase space compression,
eq \ref{eq:14} is from the heat bath variables alone.

As is evident from the examples above, it is important to identify
all the conservation laws satisfied by a given set of dynamical equations.
For instance, in many thermostatting schemes, the psuedo-Hamiltonian,
$\mathcal{H}$ is usually not the only conserved quantity. In such
cases, the probability distribution has to be sampled from the hypersurface
defined in eq \prettyref{eq:19}\cite{key-8}. We note here that all
conserved quantities have to be time reversal invariant for the present
proof to go through.

\subsection{GAUSSIAN ISOKINETIC ENSEMBLE}

Another example where the non-Hamiltonian phase space formalism can
be readily applied is the case of Gaussian isokinetic ensemble, which
keeps the kinetic energy of the system constrained to a particular
value but generates a canonical distribution in the coordinate space.
The equations of motion read

\begin{equation}
\mathbf{\dot{r}_{i}=}\dfrac{\mathbf{p_{i}}}{m_{i}}\quad i=1,..,N\label{eq:75}\end{equation}

\begin{flushleft}
\begin{equation}
\mathbf{\dot{p_{i}}=F_{i}-\left[\frac{\overset{N}{\underset{j=1}{\sum}}F_{j}.p_{j}/\mathbf{m_{j}}}{\overset{N}{\underset{j=1}{\sum}}p_{j}^{2}/\mathbf{m_{j}}}\right]}\mathbf{p_{i}}\quad i=1,..,N\label{eq:76}\end{equation}
 This isokinetic ensemble method different from other non-Hamiltonian
phase space methods in that there are no extra degrees of freedom
appended to the system, and also the total energy of the system is
also not conserved. But, by construction, one has the conservation
of the total kinetic energy and the unnormalized microcanonical probability
density can still be defined as: 
\par\end{flushleft}

\begin{equation}
p(\mathbf{p})=\delta(\overset{N}{\underset{i=1}{\sum}}\frac{\mathbf{p_{i}^{2}}}{m}-K)\label{eq:77}\end{equation}
 where $K$ is an arbitrary constant. The phase space compressibility
rate of this system, can be obtained as

\begin{equation}
\Lambda=\overset{N}{-\underset{i=1}{\sum}}\nabla_{\mathbf{r_{i}}}.\mathbf{\dot{r_{i}}}+\nabla_{\mathbf{p_{i}}}.\mathbf{\dot{p_{i}}}\label{eq:78}\end{equation}
 from eq \prettyref{eq:75} and eq \prettyref{eq:76} this becomes
gives

\begin{equation}
\Lambda=-\frac{3N-1}{K}\frac{d\phi(\{\mathbf{r}\})}{dt}\label{eq:79}\end{equation}
 where the assumption is that there is no explicit time dependence
of the potential on time, so that the partial derivative of the potential
is zero.

The Compressibility for this system is given as

\begin{equation}
\triangle S=-\frac{3N-1}{K}\overset{\tau}{\underset{0}{\int}}\frac{d\phi(\{\mathbf{r}\})}{dt}=-k_{B}\frac{3N-1}{K}\left[\phi(\mathbf{r}_{B})-\phi(\mathbf{r}_{A})\right]\label{eq:80}\end{equation}

The above equation implies, as the system starts from $\mathbf{z_{A}\equiv(}\mathbf{r}_{A},\mathbf{p_{A})}$
and evolves to $\mathbf{z_{B}\equiv(}\mathbf{r_{B}},\mathbf{p_{B})}$,
the entropy generated is proportional to the change in the potential
energy at the end points, $\phi(\mathbf{r}_{A})-\phi(\mathbf{r}_{B})$.
As the constraint, total kinetic energy, is invariant under time reversal
and the phase space compressibility of the system eq\prettyref{eq:80}
satisfies eq\prettyref{eq:22}, the present derivation of the detailed
fluctuation theorem applies for the system evolving under the Gaussian
isokinetic ensemble.

The absence of external degrees of freedom in this example means that
the phase space vector evolves deterministically and hence the probabilities
$P(\mathbf{z}_{B},\triangle S|\mathbf{z}_{A})$ and $P(\mathbf{z}_{B}^{*},-\triangle S|\mathbf{z}_{A}^{*})$
are reduced to just product of delta functions as there are no heat
bath variables to integrate over. The joint probability $P(\mathbf{z}_{B},\triangle S|\mathbf{z}_{A})$
is actually just a function of $\mathbf{z_{B}}$ and $\mathbf{z_{A}}$alone
as $\triangle S$ itself is a function of $\mathbf{z_{B}}$ and $\mathbf{z_{A}}$.
In that case, consider the probability $P(\mathbf{r}_{A}\rightarrowtail\mathbf{r}_{B})\equiv P(\mathbf{r}_{B}|r_{A})$
that the system evolves from a position$\mathbf{r}_{A}$ to a position
$\mathbf{r}_{B}$ in a time $\tau$. This involves integrating over
all momentum variables and all possible phase space compression values.

\begin{equation}
P(\mathbf{r}_{A}\rightarrowtail\mathbf{r}_{B})=\int d\mathbf{p_{A}}d\mathbf{p_{B}}\int P(\mathbf{\triangle}S)d(\mathbf{\triangle}S)P(\mathbf{z}_{B}\equiv(\mathbf{r}_{B},\mathbf{p_{B}}),\triangle S|\mathbf{z}_{A}\equiv(\mathbf{r}_{A},\mathbf{p_{A}}))\label{eq:81}\end{equation}
 where $P(\mathbf{\triangle}S)$ is the probability distribution of
$\mathbf{\triangle}S$. Since the phase space compression depends
only on the initial and final coordinates, eq \prettyref{eq:80},
we have \begin{equation}
P(\mathbf{\triangle}S)=\delta\left[\Delta S-\hat{\Delta S}(\mathbf{r}_{A},\mathbf{r}_{B})\right]\label{eq:82}\end{equation}
 where

\begin{equation}
\hat{\Delta S}(\mathbf{r}_{A},\mathbf{r}_{B})=-k_{B}\frac{3N-1}{K}\overset{\tau}{\underset{0}{\int}}\left(\frac{d\phi(\mathbf{r})}{dt}\right)dt\label{eq:83}\end{equation}
 Inserting this into the above equation

\begin{equation}
P(\mathbf{r}_{A}\rightarrowtail\mathbf{r}_{B})=\int d\mathbf{p_{A}}d\mathbf{p_{B}}d(\mathbf{\triangle}S)\delta\left[\Delta S-\hat{\Delta S}(\mathbf{r}_{A},\mathbf{r}_{B})\right]e^{\frac{\triangle S}{k_{B}}}P(\mathbf{z}_{A}^{*},-\triangle S|\mathbf{z}_{B}^{*})\label{eq:84}\end{equation}
 Now, $\hat{\Delta S}(\mathbf{r}_{A},\mathbf{r}_{B})=-\hat{\Delta S}(\mathbf{r}_{B},\mathbf{r}_{A})$
hence

\begin{equation}
P(\mathbf{r}_{A}\rightarrowtail\mathbf{r}_{B})=e^{\frac{\hat{\Delta S}(\mathbf{r}_{A},\mathbf{r}_{B})}{k_{B}}}\int d\mathbf{p_{A}}d\mathbf{p_{B}}\int d(\mathbf{\triangle}S)\delta\left[\Delta S+\hat{\Delta S}(\mathbf{r}_{B},\mathbf{r}_{A})\right]P(\mathbf{z}_{A}^{*},-\triangle S|\mathbf{z}_{B}^{*})\label{eq:85}\end{equation}
 The RHS can be rearranged, remembering that the delta function is
even in its arguments,

\begin{equation}
P(\mathbf{r}_{A}\rightarrowtail\mathbf{r}_{B})=e^{\frac{\hat{\Delta S}(\mathbf{r}_{A},\mathbf{r}_{B})}{k_{B}}}\int d\mathbf{(-p_{A}})d\mathbf{(-p_{B})}\int d(\mathbf{-\triangle}S)\delta\left[\Delta S-\hat{\Delta S}(\mathbf{r}_{B},\mathbf{r}_{A})\right]P(\mathbf{z}_{A}^{*},\triangle S|\mathbf{z}_{B}^{*})\label{eq:86}\end{equation}
 The integral on the right hand side can be readily identified as
the probability of system evolving from $r_{B}$ to $r_{A}$and hence

\begin{equation}
P(\mathbf{r}_{A}\rightarrowtail\mathbf{r}_{B})=e^{\frac{\hat{\Delta S}(\mathbf{r}_{A},\mathbf{r}_{B})}{k_{B}}}P(\mathbf{r}_{B}\rightarrowtail\mathbf{r}_{A})\label{eq:87}\end{equation}
 Now, from eq\prettyref{eq:83} \begin{equation}
\hat{\Delta S}(\mathbf{r}_{A},\mathbf{r}_{B})=-k_{B}\frac{3N-1}{K}\left[\phi(\mathbf{r_{B}})-\phi(\mathbf{r_{A}})\right]\label{eq:88}\end{equation}
 Choosing the arbitrary constant $K=(3N-1)k_{B}T$, where T is the
desired temperature, we have

\begin{equation}
P(\mathbf{r}_{A}\rightarrowtail\mathbf{r}_{B})=e^{\frac{\phi(\mathbf{r_{A}})-\phi(\mathbf{r_{B}})}{k_{B}T}}P(\mathbf{r}_{B}\rightarrowtail\mathbf{r}_{A})\label{eq:89}\end{equation}

Rearranging the above equation, we see that the detailed fluctuation
theorem just boils down to the detailed balance equation in the configuration
space:\begin{equation}
e^{-\frac{\phi(\mathbf{r_{A})}}{k_{B}T}}P(\mathbf{r}_{A}\rightarrowtail\mathbf{r}_{B})=e^{-\frac{\phi(\mathbf{r_{B})}}{k_{B}T}}P(\mathbf{r}_{B}\rightarrowtail\mathbf{r}_{A})\label{eq:90}\end{equation}

The choice $K=(3N-1)k_{B}T$ is natural from the equipartitioning
theorem which says that every independent momentum degree of freedom
has an average kinetic energy of $\frac{1}{2}k_{B}T$. The system
has $(3N-1)$ independent momentum degrees of freedom ($3N$ momentum
variables and one constraint on the total kinetic energy), hence the
total kinetic energy is $\frac{1}{2}(3N-1)k_{B}T$. The Gaussian Isokinetic
ensemble is a simple yet powerful example to appreciate that the Non-Hamiltonian
characteristic alone is sufficient to bring out all the seemingly
counter-intuitive features of the statistical mechanics like the entropy
production and consumption anisotropy, direction of time emerging
from the time reversible dynamics, anisotropy in the transition probabilities,
emergence of Boltzmann distribution in the configuration space etc.

\section{DETAILED FLUCTUATION THEOREM AND CONSERVATION LAWS}

Consider the Nose Hoover thermostatting scheme. If the forces acting
on the system are derivable from a two body potential,

\begin{equation}
\Phi(\mathbf{r}_{1},...,\mathbf{r}_{N})=\overset{N}{\underset{\overset{i,j=1}{i\neq j}}{\frac{1}{2}\sum}}\phi_{2}(|\mathbf{r}_{i}-\mathbf{r}_{j}|)\label{eq:91}\end{equation}
 such that the net force acting on the system is zero, then there
are $d$ additional conserved quantities that emerge in the Nose-Hoover
dynamics:

\begin{equation}
\mathbf{P}e^{\eta}=\mathbf{K}\label{eq:92}\end{equation}
 where $\mathbf{P}=\overset{n}{\underset{i=1}{\sum}}\mathbf{p}_{i}$
is the total momentum of the system and $\mathbf{K}$ is an arbitrary
vector in $d$ dimensions. In our context, this would mean that we
can no longer use eqs \prettyref{eq:20} but instead use $p(\mathbf{\Gamma})=\delta(H^{'}(\mathbf{\Gamma})-E)\delta(\mathbf{P}e^{\eta}-\mathbf{K})$
as the correct distribution function. The problems associated with
the presence of additional conservation laws in the context of generating
the dynamics appropriate to a desired ensemble is well studied \cite{key-8},
where it is shown that the presence of hidden conservation laws in
the dynamics will lead to an ensemble different from the required
canonical ensemble. It should be noted that the suggested solution
of appending an extended chain of thermostats, instead of one, though
known to generate the correct distribution in the system subspace,
will not invalidate the presence of additional conservation laws.
So, the augmentation of the Nose-Hoover to Nose-Hoover chain will
not help in demonstration of the detailed fluctuation theorem. The
problem arises from the fact $\mathbf{P}e^{\eta}$ is not invariant
under time reversal: $(\mathbf{P}e^{\eta})^{*}\neq\mathbf{P}e^{\eta}$
and it is because of this problem, the present method of derivation
hits a roadblock. In fact, for all systems which have conserved quantities
which do not have a definite parity or are of odd parity under time
reversal would fail to capture fluctuation theorems of the usual type,
as the time reversed states are not accessible to the system. But
for systems which have conserved quantities which are of odd parity
under time reversal, say$\mathbf{K}$, it is easy to see that the
Detailed Fluctuation Theorem takes the form:

\begin{equation}
P_{+}(\mathbf{z}_{B},\triangle S|z_{A},\mathbf{K})=e^{\frac{\Delta S}{k_{B}}}P_{-}(\mathbf{z}_{B}^{*},-\triangle S|\mathbf{z}_{A}^{*},\mathbf{K^{*})}\label{eq:93}\end{equation}

\section{Free Energy Relations from the Detailed Fluctuation theorem}

It would be worthwhile to investigate whether the free energy equality,
Jarzynski's identity, is derivable from the detailed fluctuation theorem
in the non-Hamiltonian framework. For simplicity, we shall attempt
to derive the Jarzynski's identity from the detailed fluctuation theorem
result for a system coupled to Nose-Hoover thermostat. The Jarzynski's
identity \textbf{\cite{key-21}} reads,

\begin{equation}
<exp(-\beta W)>=exp(-\beta\triangle F_{AB}),\qquad\beta=\frac{1}{kT}\label{eq:94}\end{equation}
 where $\triangle F_{AB}$is the equilibrium free energy difference
between A and B:

\begin{equation}
\triangle F_{AB}=-k_{B}Tln\left(\frac{Z_{A}}{Z_{B}}\right)\label{eq:95}\end{equation}
 where $Z_{A}$and $Z_{B}$ are canonical partition functions of the
systems A and B.

\begin{equation}
Z_{A}=\int e^{-\beta H_{A}^{s}(\mathbf{r_{\mathbf{A}},}\mathbf{p_{\mathbf{A}})}}d\mathbf{r_{A}}d\mathbf{p_{A}}\label{eq:96}\end{equation}
 and similarly

\begin{equation}
Z_{B}=\int e^{-\beta H_{B}^{s}(\mathbf{r_{B},}\mathbf{p_{B})}}d\mathbf{r_{B}}d\mathbf{p_{B}}\label{eq:97}\end{equation}
 where $H_{A}^{s}$and $H_{B}^{s}$ are the Hamiltonian of the two
systems whose free energy difference is to be computed. The Hamiltonians
are superscripted to make the distinction between the Hamiltonian
of the system of interest and the psuedo-Hamiltonian, which contains
reservoir degrees of freedom also. We assume that there is a single
time-dependent Hamiltonian, which at the time $t=0$ is the Hamiltonian
corresponding to the state A, $H_{A}$ and at a time $t=\tau$ transforms
to the Hamiltonian corresponding to the state B, $H_{B}$. This variation
can be brought about by, for example, a time dependent potential $\Phi(\mathbf{r}_{1},...,\mathbf{r}_{N},t)$,
such at $\Phi(\mathbf{r}_{1},...,\mathbf{r}_{N},0)=\Phi_{A}(\mathbf{r}_{1},...,\mathbf{r}_{N})$
and $\Phi(\mathbf{r}_{1},...,\mathbf{r}_{N},\tau)=\Phi_{B}(\mathbf{r}_{1},...,\mathbf{r}_{N})$
where $\Phi_{A}$and $\Phi_{B}$ are the potentials of states A and
B respectively.

The change in the energy of the {}``system of interest'' due to
this time variation of the potential is given by \cite{key-2,key-10}:

\begin{equation}
H_{B}-H_{A}=H^{S}(\mathbf{z}_{B},\tau)-H^{S}(\mathbf{z}_{A},0)=\overset{\tau}{\underset{0}{\int}}dt\frac{\partial H^{s}(\mathbf{z}_{t},t)}{\partial\mathbf{z}_{t}}\dot{\mathbf{z}_{t}}+\overset{\tau}{\underset{0}{\int}}dt\frac{\partial H^{s}(\mathbf{z}_{t},t)}{\partial t}=Q+W\label{eq:98}\end{equation}
 The above equation can be called the mathematical formulation of
the First law of Thermodynamics, where the term on the left hand side
is identified with the change in the internal energy of the system,
the first term of the right is identified as the heat Q transferred
from the bath to the system and the second term is the work performed
on the the system. Note that both, work and heat are defined in terms
of the system Hamiltonian alone. The effect of thermostatting is felt
only through the coupling of evolution of the system variables and
heat bath variables.

From the explicit functional form for the system Hamiltonian and the
Nose-Hoover equations of motion, we can calculate the first term on
the right hand side of eq \prettyref{eq:98} as

\begin{equation}
Q=\overset{\tau}{\underset{0}{\int}}dt\left(-\dfrac{p_{\eta}(t)}{Q_{1}}\right)\overset{N}{\underset{i=1}{\sum}}\frac{\mathbf{p_{i}}^{2}(t)}{m}\label{eq:99}\end{equation}
 Identifying the term $\dfrac{p_{\eta}(t)}{Q_{1}}$ as $\dot{\eta}$
and $\overset{N}{\underset{i=1}{\sum}}\frac{\mathbf{p_{i}}^{2}(t)}{m}$
as $\dot{p_{\eta}}+dNk_{B}T$ from the Nose-Hoover equations of motion
eq (38) and eq (39) we get

\begin{equation}
Q=-\overset{\tau}{\underset{0}{\int}\dot{\eta}}(\dot{p_{\eta}}+dNk_{B}T)dt\label{eq:100}\end{equation}
 As discussed earlier, If the times $t=0$ and $t=\tau$ are such
long after the system has reached steady state such that the average
kinetic energy is determined by the temperature, we have, from eq\prettyref{eq:38}
and eq\prettyref{eq:53}

\begin{equation}
Q(\tau)=-dNk_{B}T\overset{\tau}{\underset{0}{\int}}dt\dot{\eta}(t)=-dNk_{B}T(\eta(\tau)-\eta(0))\label{eq:101}\end{equation}
 Identifying $dN\dot{\eta}(t)$ as the phase space compression rate
for this system, eq\prettyref{eq:43}, we have \begin{equation}
Q(\tau)=k_{B}T\overset{\tau}{\underset{0}{\int}}\Lambda(t)dt=-T\triangle S\label{eq:102}\end{equation}
 Thus we see that $k_{B}T$ times the total phase space compression
can be identified with the heat lost by the system to the thermostat.
This is consistent with the assumption of eq \prettyref{eq:3}. If
the term $\triangle S$ can be identified by the change in the entropy,
then the above equation boils to the Second Law of Thermodynamics,
\cite{key-19,key-20}

With this identification, we are ready to take-on the Jarzynski's
work theorem in the context of Nose-Hoover thermostatted system.

\begin{equation}
<e^{(-\beta W)}>=\int d\mathbf{z}_{A}p(\mathbf{z}_{A})P_{+}(\mathbf{z}_{B},\triangle S|\mathbf{z}_{A})e^{(-\beta W(\mathbf{z_{A},z_{B},\triangle S)}))}d(\mathbf{z_{B}})d(\triangle S)\label{eq:103}\end{equation}
 It is easy to understand how the above integral is constructed. the
quantity $P_{+}(\mathbf{z}_{B},\triangle S|\mathbf{z}_{A})$ gives
the probability of system making a transition from $\mathbf{z}_{A}$
at time $t=0$ to $\mathbf{z}_{B}$ at time $t=\tau$ and in the process
generates a phase space compression of $\triangle S$. $p(\mathbf{z}_{A})$
denotes the probability that the system is found in the state $\mathbf{z}_{A}$at
time $t=0.$ If the times are so chosen that the system is fully equilibrated
at time t=0, this probability is actually the probability of the canonical
ensemble:

\begin{equation}
p(\mathbf{z}_{A})=\frac{e^{-\beta H_{A}^{s}(\mathbf{Z_{A})}}}{\mathbf{Z}_{A}}\label{eq:104}\end{equation}
 where $\mathbf{Z}_{A}$ is the partition function for the system
in the state A, eq \prettyref{eq:96}. Substituting this and also
the detailed fluctuation theorem, we have

\begin{equation}
<e^{-\beta W}>=\frac{1}{\mathbf{Z}_{A}}\int d\mathbf{z}_{A}e^{-\beta H_{A}^{s}(\mathbf{Z_{A})}}e^{\frac{\triangle S}{k_{B}}}P(\mathbf{z}_{A}^{*},-\triangle S|\mathbf{z}_{B}^{*})e^{(-\beta W(\mathbf{z_{A},z_{B},\triangle S)}))}d(\mathbf{z_{B}})d(\triangle S)\label{eq:105}\end{equation}
 Consider the terms in the exponentials, from eq\prettyref{eq:105}
and eq \prettyref{eq:98} we have \begin{equation}
-\beta\left[H_{A}^{s}(\mathbf{z}_{A})-T\triangle S+W(\mathbf{z_{A},z_{B},\triangle}S)\right]=-\beta(H_{A}^{s}(\mathbf{z}_{B})\label{eq:106}\end{equation}
 and thus

\begin{equation}
<e^{-\beta W}>=\frac{1}{\mathbf{Z}_{A}}\int d\mathbf{z}_{A}e^{-\beta H_{B}^{s}(\mathbf{z_{B})}}P(\mathbf{z}_{A}^{*},-\triangle S|\mathbf{z}_{B}^{*})d(\mathbf{z_{B}})d(\triangle S)\label{eq:107}\end{equation}
 But \begin{equation}
\int P(\mathbf{z}_{A}^{*},-\triangle S|\mathbf{z}_{B}^{*})d\mathbf{z}_{A}d(\triangle S)=1\label{eq:108}\end{equation}
 And hence we have

\begin{equation}
<e^{-\beta W}>=\frac{1}{\mathbf{Z}_{A}}\int d\mathbf{z}_{B}e^{-\beta H_{B}^{s}(\mathbf{z_{B})}}=\frac{\mathbf{Z}_{B}}{\mathbf{Z}_{A}}=exp(-\beta\triangle F_{AB})\label{eq:109}\end{equation}

This is the Jarzynski's Work theorem, eq\prettyref{eq:95} we set
out to prove.

Although the connection between the detailed Fluctuation theorem and
the Jarzynski's identity has been established here only for the case
of Nose-Hoover thermostatting scheme, it should be evident that the
Jarzynski's work theorem can be derived in all contexts where the
Detailed Fluctuation theorem is applicable.

For illustration, consider the case of Gaussian Isokinetic ensemble.
As mentioned above, these equations of motion fail to generate the
proper canonical sampling in the momentum space but generates a canonical
distribution in the coordinate space. But from eq, we see that the
Free energy differences depends on the logarithm of the ratio of the
two partition functions and hence the momentum partition function
cancels out in the ratio and we are left with the ratio of the configuration
partition functions at the states A and B. So we can see that one
can realize the Jarzynski's identity in the Gaussian Isokinetic ensemble.

As with the Nose-Hoover thermostat example above, Consider a system
given by the Hamiltonian, $H^{s}(\mathbf{\{r}\},\{\mathbf{p\},}t)=\overset{N}{\underset{i=1}{\sum}}\dfrac{\mathbf{p}_{i}^{2}}{2m_{i}}+\Phi(\{\mathbf{r\}},t)$
we have the Gaussian Isokinetic equations of motion of the form,

\begin{equation}
\mathbf{\dot{r}_{i}=}\dfrac{\mathbf{p_{i}}}{m_{i}}\quad i=1,..,N\label{eq:110}\end{equation}

\begin{center}
\begin{equation}
\mathbf{\dot{p_{i}}=-\nabla_{\mathbf{r}_{j}}\phi(\{\mathbf{r}\},t)}+\mathbf{\alpha(}\{\mathbf{r\}},\mathbf{\{p}\},t)\mathbf{p_{i}}\quad i=1,..,N\label{eq:111}\end{equation}

\par\end{center}

where \begin{equation}
\mathbf{\alpha(}\{\mathbf{r\}},\mathbf{\{p}\},t)=\left[\frac{\overset{N}{\underset{j=1}{\sum}}\nabla_{\mathbf{r}_{j}}\phi(\{\mathbf{r}\},t).p_{j}/\mathbf{m_{j}}}{\overset{N}{\underset{j=1}{\sum}}p_{j}^{2}/\mathbf{m_{j}}}\right]\label{eq:112}\end{equation}

and the potential $\phi(\{\mathbf{r}\},t)$ is such that $\Phi(\mathbf{r}_{1},...,\mathbf{r}_{N},0)=\Phi_{A}(\mathbf{r}_{1},...,\mathbf{r}_{N})$
and $\Phi(\mathbf{r}_{1},...,\mathbf{r}_{N},\tau)=\Phi_{B}(\mathbf{r}_{1},...,\mathbf{r}_{N})$
where $\Phi_{A}$and $\Phi_{B}$ are the potentials of states A and
B respectively. The time variation of this potential indicates that
the work is done on the system. Consider again, eq\prettyref{eq:98},
we have the heat lost to the thermostat given by

\begin{equation}
Q=\overset{\tau}{\underset{0}{\int}}dt\frac{\partial H^{s}(\mathbf{z}_{t},t)}{\partial\mathbf{z}_{t}}\dot{\mathbf{z}_{t}}=\overset{\tau}{\underset{0}{\int}}dt\mathbf{\alpha(}\{\mathbf{r\}},\mathbf{\{p}\},t)\overset{N}{\underset{i=1}{\sum}}p_{i}^{2}/\mathbf{m_{i}}\label{eq:113}\end{equation}
 substituting for $\mathbf{\alpha(}\{\mathbf{r\}},\mathbf{\{p}\},t)$
and simplifying we have \begin{equation}
Q=\overset{\tau}{\underset{0}{\int}}dt\left[\overset{N}{\underset{j=1}{\sum}}\nabla_{\mathbf{r}_{j}}\phi(\{\mathbf{r}\},t).p_{j}/\mathbf{m_{j}}\right]\label{eq:114}\end{equation}
 Identifying $\overset{N}{\underset{j=1}{\sum}}\nabla_{\mathbf{r}_{j}}\phi(\{\mathbf{r}\},t).p_{j}/\mathbf{m_{j}}=\frac{K}{3N-1}\Lambda(t)$
we have $Q=\frac{K}{3N-1}\overset{\tau}{\underset{0}{\int}}dt\Lambda(t)$

Together with eq \prettyref{eq:44} this gives \begin{equation}
Q=\frac{-K}{3N-1}\frac{\Delta S}{k_{B}}\label{eq:115}\end{equation}
 Choosing the arbitrary constant $K=(3N-1)k_{B}T$ (for reasons already
discussed ) we have

\begin{equation}
Q=-T\Delta S\label{eq:116}\end{equation}
 Again, we see that the heat lost by the system is proportional to
the total phase space compression, as with the Nose-Hoover thermostat.
With this identification, the procedure to calculate $<e^{-\beta W}>$
is essentially unchanged from the Nose-Hoover case, and we have the
Jarzynski's identity in the case of a system coupled to a Gaussian
Isokinetic ensemble.

\section{Conclusion}

Detailed Fluctuation Theorem has been extended to a class of thermostatted
systems, evolving under the extended system dynamics. It is demonstrated
that this theorem retains the same form as for the original DFT for
entropy production when one replaces the thermodynamic entropy with
phase space compressibility. This theorem is of a wider applicability
than its original counterpart and can be applied even to the systems
at equilibrium. It is shown that this detailed fluctuation theorem
is formally equivalent to the detailed balance equation for systems
at equilibrium. Rederivation of the Jarzynski's identity through the
Detailed Fluctuation theorem has been demonstrated for both Nose-Hoover
thermostat and the Gaussian Isokinetic ensembles. 

\medskip{}


\begin{thebibliography}{18}
\bibitem[1]{key-6}Jarzynski C, Hamiltonian Derivation of a Detailed
Fluctuation Theorem: arXiv:cond-mat/9908286v1

\bibitem[2]{key-7}Jarzynski C, Hamiltonian Derivation of a Detailed
Fluctuation Theorem, Journal of Statistical Physics, Volume \textbf{98},
Numbers 1-2 / January, 2000, DOI: 10.1023/A:1018670721277

\bibitem[3]{key-8}Mark E. Tuckerman, Yi Liu, Giovanni Ciccotti, and
Glenn J. Martyna, Non-Hamiltonian molecular dynamics: Generalizing
Hamiltonian phase space principles to non-Hamiltonian systems J. Chem.
Phys. \textbf{115}, 1678 (2001)

\bibitem[4]{key-9}Alessandro Sergi and Mauro Ferrario, Non-Hamiltonian
equations of motion with a conserved energy, Phys. Rev. E \textbf{64},
056125 (2001)

\bibitem[5]{key-14}EGD Cohen, L. Rondoni Note on Phase Space Contraction
and Entropy Production in Thermostatted Hamiltonian Systems, arXiv:cond-mat/9712213v1,

\bibitem[6]{key-13}Giovanni Galavotti Fluctuation relation, fluctuation
theorem, thermostats and entropy creation in nonequilibrium statistical
physics, C. R. Physique \textbf{8 }(2007) 486\textendash{}494.

\bibitem[7]{key-12}D. Daems, G. Nicolis Entropy production and phase
space volume contraction, Phys Rev E \textbf{59}, 4000 (1999)

\bibitem[8]{key-11}DJ Evans, DJ Searles - The fluctuation theorem-Advances
in Physics, 2002, Vol. \textbf{51}, No. 7, 1529-1585

\bibitem[9]{key-10}Michel A. Cuendet, Statistical Mechanical Derivation
of Jarzynski's Identity for Thermostatted Non-Hamiltonian Dynamics.
Phys. Rev. Lett. \textbf{96}, 120602 (2006).

\bibitem[10]{key-1}E. Schöll-Paschinger and C. Dellago, A proof of
Jarzynski's nonequilibrium work theorem for dynamical systems that
conserve the canonical distribution J. Chem. Phys \textbf{125}, 054105
2006

\bibitem[11]{key-2}The Jarzynski identity derived from general Hamiltonian
or non-Hamiltonian dynamics reproducing NVT or NPT ensembles Michel
A. Cuendet, J. Chem Phys, \textbf{125}, 144109 2006

\bibitem[12]{key-3}Riccardo Chelli,a Simone Marsili, Alessandro Barducci,
and Piero Procacci, Recovering the Crooks equation for dynamical systems
in the isothermal-isobaric ensemble: a strategy based on the equations
of motion. J. Chem. Phys \textbf{126}, 044502 (2007)

\bibitem[13]{key-15}Riccardo Chelli, Simone Marsili, Alessandro Barducci,
and Piero Procacci, Generalization of the Jarzynski and Crooks nonequilibrium
work theorems in molecular dynamics simulations, Phys Rev E, \textbf{75},
050101R 2007

\bibitem[14]{key-16}M. E. Tuckerman, C. J. Mundy and G. J. Martyna,
On the classical statistical mechanics of non-Hamiltonian systems,
Euro Phys Lett, \textbf{45} (2), pp. 149-155 (1999)

\bibitem[15]{key-18}Statistical Mechanics: Theory and Molecular Simulation
Mark. E. Tuckerman, Oxford Graduate Texts.

\bibitem[16]{key-19}Williams SR, Searles DJ, Evans DJ. 2004. Independence
of the transient fluctuation theorem to thermostatting details. Phys.
Rev. E \textbf{70} 066113

\bibitem[17]{key-20}Bright JN, Evans DJ, Searles DJ. 2005. New observations
regarding deterministic, time-reversible thermostats and Gauss's principle
of least constraint. J. Chem. Phys. \textbf{122} 194106

\bibitem[18]{key-21}C. Jarzynski Nonequilibrium Equality for Free
Energy Differences, Phys. Rev. Lett. \textbf{78}, 2690\textendash{}2693
(1997)

\end{thebibliography}
\end{document}